\documentclass[runningheads]{llncs}
\usepackage[T1]{fontenc}
\usepackage{graphicx}
\usepackage{amsmath,amssymb,amsfonts}
\usepackage{algorithm}
\usepackage{algorithmic}
\usepackage{textcomp}
\begin{document}
\title{Application of Hybrid Chain Storage Framework in Energy Trading and Carbon Asset Management}
\titlerunning{Hybrid Chain Storage for Energy Trading}
\author{Yinghan Hou\inst{1}\thanks{Yinghan Hou and Zongyou Yang contributed equally to this work.} \and
Zongyou Yang\inst{2}$^{\star}$ \and
Xiaokun Yang\inst{3}\thanks{Corresponding author.}}
\authorrunning{Y. Hou et al.}
\institute{Department of Earth Science and Engineering, Imperial College London, London, United Kingdom\\
\email{ghwzhyinghan@gmail.com} \and
Department of Computer Science, University College London, London, United Kingdom\\
\email{dryang0624@gmail.com} \and
School of Electronic Information, Nanchang Institute of Technology, Nanchang, China\\
\email{yangxk@bupt.cn}}
\maketitle              % typeset the header of the contribution
\begin{abstract}
Distributed energy trading and carbon asset management involve high-frequency, small-value settlements with strong audit requirements. Fully on-chain designs incur excessive cost, while purely off-chain approaches lack verifiable consistency. This paper presents a hybrid on-chain and off-chain settlement framework that anchors settlement commitments and key constraints on-chain and links off-chain records through deterministic digests and replayable auditing. Experiments under publicly constrained workloads show that the framework significantly reduces on-chain execution and storage cost while preserving audit trustworthiness.

\keywords{Blockchain \and Distributed Energy Trading \and Carbon Asset Management \and Hybrid On/Off-Chain Storage}
\end{abstract}
\section{Introduction}

Distributed energy trading and carbon asset management are characterized by high-frequency, small-value settlements under strict regulatory and audit requirements. These systems must support consistency verification, post hoc auditing, and controlled long-term cost. Blockchain offers tamper resistance and public verifiability, yet fully on-chain execution incurs prohibitive cost under high-frequency workloads. Purely off-chain approaches reduce cost but lack independently verifiable consistency \cite{eyal_bitcoinng,aitzhan_energy_security,zyskind_privacy}. Existing work tends to trade cost efficiency for audit trustworthiness, a tension that becomes particularly pronounced in high-frequency and strongly regulated scenarios \cite{saxena_energy_field,vishwakarma_p2p_energy,boumaiza_energy_carbon}.

To address this gap, this paper proposes a hybrid on-chain and off-chain framework that anchors commitments and critical constraints on-chain while retaining settlement details off-chain. Deterministic settlement digests enable replayable auditing without additional trust assumptions. For carbon asset management, lifecycle quantity conservation is enforced as an on-chain invariant without replacing official verification. Experiments demonstrate that the framework significantly reduces on-chain cost while preserving auditability \cite{zhang_carbon_shanghai,schneider_integrity}.

\section{Related Work}

\subsection{Blockchain for Distributed Energy Trading}

Blockchain has been applied to distributed energy trading to reduce trust costs and provide verifiable records. Saxena et al.\ demonstrated feasibility through field deployment but did not address long-term costs under high-frequency settlement \cite{saxena_energy_field}. Aitzhan and Svetinovic focused on security and privacy, without supporting replayable third-party auditing \cite{aitzhan_energy_security}. Vishwakarma et al.\ incorporated loss traceability while largely assuming on-chain settlement, with limited cost analysis \cite{vishwakarma_p2p_energy}. Liu et al.\ proposed multi-modal modeling for automating smart contract generation, demonstrating how structured design can reduce manual coding effort in blockchain applications \cite{liu_smart_contract}. Stepanova and Eri\c{n}\v{s} presented a blockchain-based model for professional growth data processing, highlighting the applicability of blockchain to structured record management and verifiable data workflows \cite{stepanova_blockchain_growth}.

\subsection{Blockchain for Carbon Asset Management and Integrity Constraints}

In carbon asset management, blockchain is mainly used as a registration and record layer. Zhang et al.\ highlighted traceability benefits alongside on-chain cost and regulatory constraints \cite{zhang_carbon_shanghai}. Schneider et al.\ showed that environmental integrity and double counting are structural issues beyond purely technical solutions \cite{schneider_integrity}. This motivates enforcing only minimal consistency constraints on-chain.

\subsection{Privacy-Preserving Validation and Commitments}

Privacy-preserving approaches combine off-chain data with on-chain verification. Zyskind et al.\ focused on access control rather than replayable auditing \cite{zyskind_privacy}. Zero-knowledge tools such as Bulletproofs incur non-trivial overhead under high-frequency workloads \cite{bunz_bulletproofs}. Dynamic accumulators support membership verification, but cost stability is less explored \cite{camenisch_accumulator}.

\subsection{Gap and Positioning of This Work}

Prior work lacks system-level solutions for high-frequency, small-value, and audit-intensive scenarios. Fully on-chain designs face cost pressure, while fully off-chain designs lack independent verification. This work addresses settlement and audit infrastructure by deterministically linking on-chain commitments with off-chain records to enable replayable auditing at controlled cost.

\section{Problem Statement and Design Goals}

\subsection{Scenario Definition}

This work studies peer-to-peer inspired distributed energy trading and carbon asset management scenarios. Energy prosumers participate in decentralized trading and settlement under regulatory and audit requirements. Grid operators handle physical transmission only. Carbon authorities issue and verify assets. Auditors independently verify settlement consistency and asset conservation ex post. These scenarios are characterized by high-frequency arrivals, small-value transactions, and strong audit constraints, without a single trusted authority. This work does not model market clearing or trading games, and focuses solely on settlement and audit infrastructure to achieve verifiable consistency with controlled cost.

\subsection{Assumptions and Goals}

We adopt an honest-but-curious threat model \cite{zyskind_privacy}. The on-chain execution environment is assumed to be tamper-resistant and publicly verifiable. Off-chain data are not trusted and are validated only through consistency with on-chain commitments. Auditors are independent third parties relying solely on public on-chain state and obtainable off-chain inputs.

Under these assumptions, the system design has three goals. First, it should support replayable auditing so that third parties can independently verify consistency between off-chain records and on-chain commitments. Second, it should remain cost-efficient under high-frequency workloads by minimizing on-chain state and execution \cite{eyal_bitcoinng,poon_lightning}. Third, it should enable privacy-friendly verification by avoiding mandatory disclosure of complete transaction or asset information \cite{bunz_bulletproofs,camenisch_accumulator}.

\section{System Overview}

The proposed framework stores full transaction details off-chain while retaining only commitments and minimal critical states on-chain. The on-chain layer enforces mandatory constraints, and the off-chain layer performs deterministic digest construction and optional batch compression. The central audit objective is reproducibility: off-chain records are mapped to on-chain commitments through public and deterministic rules, enabling auditors to independently recompute and verify results.

\subsection{System Components and Boundaries}

\begin{figure*}[t]
\centering
\includegraphics[width=\textwidth]{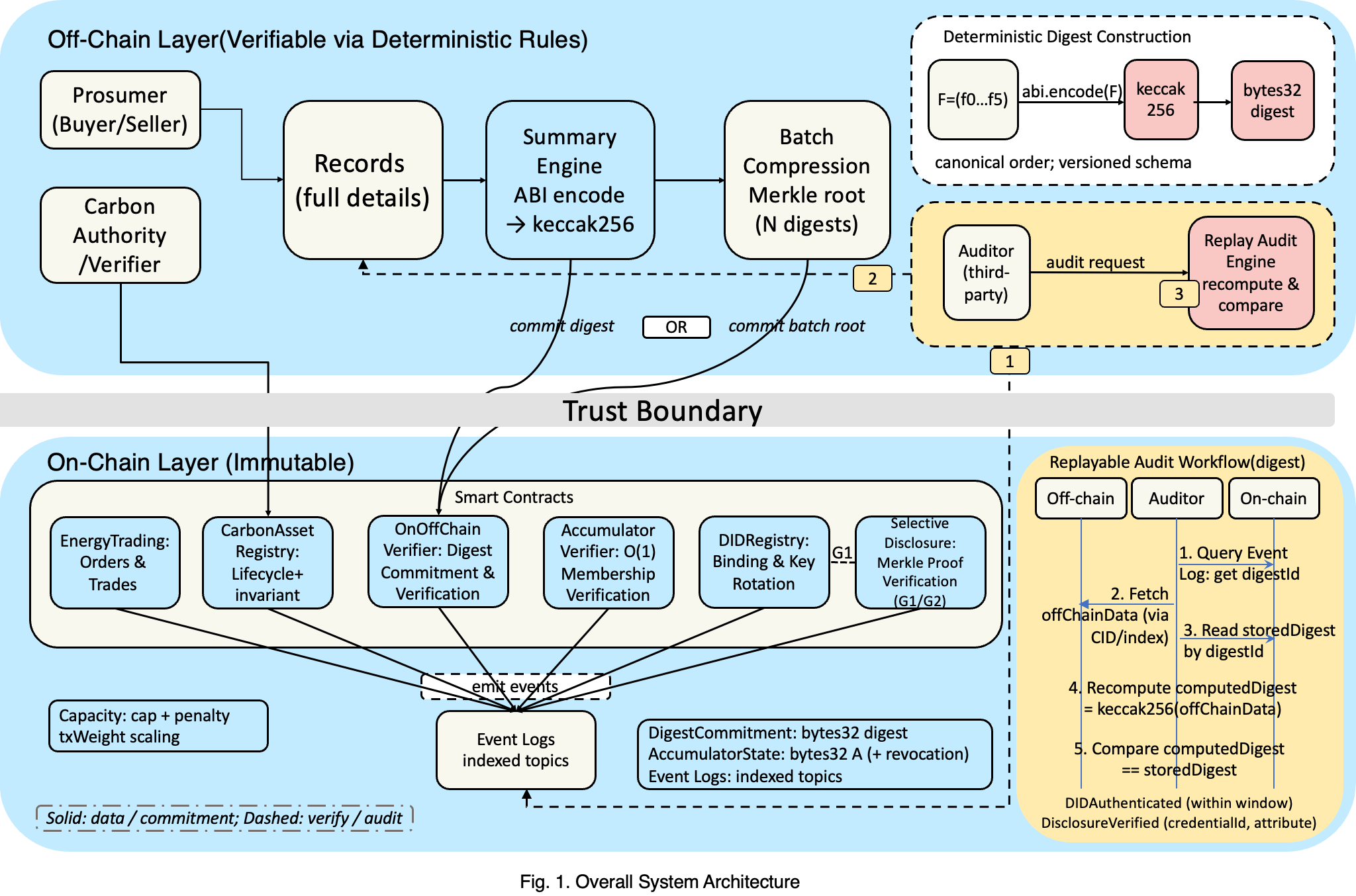}
\caption{Overall system architecture and trust boundary. The off-chain layer constructs deterministic digests from full records, while the on-chain layer anchors commitments and enforces constraints.}
\label{fig:system_overview}
\end{figure*}

The system includes three participant roles. Prosumers generate orders and settlement records off-chain and submit on-chain commitments when required. Carbon authorities execute on-chain operations for asset registration and verification. Auditors read commitments from the blockchain and replay off-chain records for verification. The on-chain layer stores only commitments, aggregated states, and traceable events, while the off-chain layer stores full records and performs digest construction. Fig.~\ref{fig:system_overview} illustrates the architecture.

\subsection{Audit Core Formulation and Complexity}

For each settlement record, the system adopts a fixed and ordered set of fields. The off-chain layer ABI (Application Binary Interface) encodes the fields and computes the digest.
\begin{equation}
\mathit{digest} = \mathit{keccak256}(\mathit{offChainData})
\label{eq:digest}
\end{equation}

The on-chain layer stores only the digest or a batch root and does not parse field semantics. Auditors recompute the digest using the same rules and compare it with the on-chain commitment.
\begin{equation}
\mathit{computedDigest} = \mathit{storedDigest}
\label{eq:digest_check}
\end{equation}

When batching is used, off-chain digests are organized into a Merkle tree and only the root is committed. The inclusion verification of a single record has complexity $O(\log n)$.

The system introduces an RSA accumulator to support constant-time membership verification. Each element is mapped to a prime representative $r_i$ (e.g., hash-to-prime) before accumulation. This verification path executes a fixed number of steps on-chain. Fig.~\ref{fig:onchain_mechanisms} illustrates the corresponding mechanism.
The accumulator value is defined as
\begin{equation}
A = g^{\prod_i r_i} \bmod N
\label{eq:accumulator}
\end{equation}
Off-chain witnesses are maintained as
\begin{equation}
W_i = g^{\prod_{j \neq i} r_j} \bmod N
\label{eq:witness}
\end{equation}
On-chain verification evaluates
\begin{equation}
W_i^{r_i} \bmod N \stackrel{?}{=} A
\label{eq:accumulator_verify}
\end{equation}

Algorithm~\ref{alg:settlement_audit} summarizes the end-to-end settlement and audit workflow, covering off-chain digest construction, on-chain commitment anchoring, and third-party replay verification.

\begin{algorithm}[t]
\caption{Hybrid Settlement and Audit Workflow}
\label{alg:settlement_audit}
\begin{algorithmic}[1]
\renewcommand{\algorithmicrequire}{\textbf{Phase 1: Off-chain Digest Construction}}
\REQUIRE
\STATE Collect settlement records $\{R_1, R_2, \ldots, R_n\}$
\FOR{each record $R_i$}
    \STATE $d_i \leftarrow \mathrm{keccak256}(\mathrm{ABI.encode}(R_i))$
\ENDFOR
\STATE Build Merkle tree $T$ over $\{d_1, \ldots, d_n\}$
\STATE $\mathit{root} \leftarrow T.\mathrm{root}()$
\renewcommand{\algorithmicrequire}{\textbf{Phase 2: On-chain Commitment}}
\REQUIRE
\STATE Submit $\mathit{root}$ to \texttt{OnOffChainVerifier}
\STATE Contract stores $\mathit{root}$ and emits event with batch metadata
\renewcommand{\algorithmicrequire}{\textbf{Phase 3: Auditor Replay Verification}}
\REQUIRE
\STATE Retrieve $\mathit{root}_{\mathrm{stored}}$ from on-chain event log
\STATE Obtain off-chain records $\{R_1, \ldots, R_n\}$
\FOR{each record $R_i$}
    \STATE $d_i' \leftarrow \mathrm{keccak256}(\mathrm{ABI.encode}(R_i))$
\ENDFOR
\STATE $\mathit{root}' \leftarrow \mathrm{MerkleTree}(\{d_1', \ldots, d_n'\}).\mathrm{root}()$
\IF{$\mathit{root}' = \mathit{root}_{\mathrm{stored}}$}
    \STATE \textbf{return} \textsc{Consistent}
\ELSE
    \STATE \textbf{return} \textsc{Mismatch Detected}
\ENDIF
\end{algorithmic}
\end{algorithm}

\subsection{Carbon Asset Consistency}

To address the challenge of verifying carbon asset integrity, we implement a mechanism of \textit{On-chain Algorithmic Regulation}. The system enforces lifecycle conservation constraints directly on-chain. The contract maintains aggregated states and checks the upper bound of available balances. The core invariant is
\begin{equation}
\mathit{availableCredits} + \mathit{retiredCredits} \le \mathit{totalCredits}
\label{eq:carbon_conservation}
\end{equation}
For transfer and retirement operations, the contract also enforces
\begin{equation}
\mathit{amount} \le \mathit{availableCredits}
\label{eq:transfer_bound}
\end{equation}
Requests that violate these conditions are reverted, which prevents the on-chain state from entering irreversible inconsistency. The asset component in Fig.~\ref{fig:onchain_mechanisms} corresponds to this constraint.

\subsection{Two-Layer Gated Selective Disclosure}

\begin{figure*}[t]
\centering
\includegraphics[width=\textwidth]{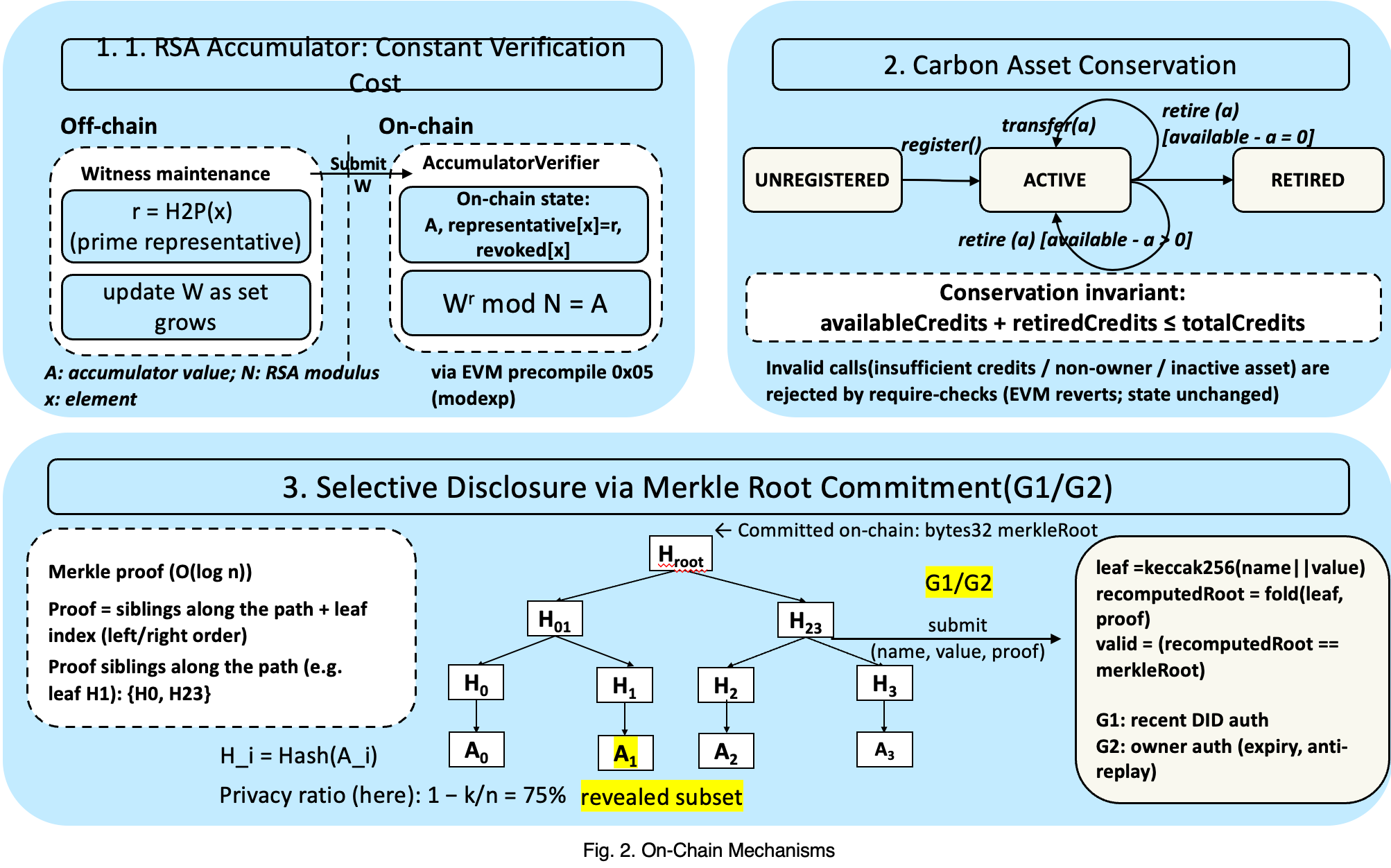}
\caption{Core on-chain mechanisms. (a) Constant-time RSA accumulator verification with off-chain witness maintenance. (b) Carbon asset lifecycle conservation enforced by balance invariants. (c) Selective disclosure based on Merkle root commitments with identity-gated access control.}
\label{fig:onchain_mechanisms}
\end{figure*}

The system adopts a two-layer gated selective disclosure mechanism. The first layer performs identity authentication within a valid window; unauthenticated requests cannot proceed. The second layer enforces attribute-level gating: the requester submits an authorization signature and corresponding Merkle proof, and the contract verifies both before returning the result. The disclosure component in Fig.~\ref{fig:onchain_mechanisms} corresponds to this workflow.

\subsection{Trust Model}

The system adopts an honest-but-curious threat model \cite{zyskind_privacy}, where participants follow the protocol but may attempt to infer additional information. On-chain states are immutable and serve as stable trust anchors, while off-chain data are validated only through consistency with on-chain commitments via deterministic digest recomputation. Audit conclusions are derived solely from public rules without relying on private keys or trusted hardware.

\section{Smart Contract Design}

The system consists of six functionally decoupled smart contracts. The design follows the principle of minimal necessary on-chain state. Only commitment anchors, aggregated states, and constraint checks are retained on-chain. High-frequency details remain off-chain and are linked through commitments and events.

\textbf{OnOffChainVerifier} anchors settlement digests or batch roots generated off-chain without parsing business fields. Settlement consistency is established by auditor recomputation.
\textbf{EnergyTrading} maintains orders and minimal settlement fields for matching and state updates; high-frequency details are linked through off-chain records and on-chain events.
\textbf{CarbonAssetRegistry} manages registration, transfer, and retirement on-chain, enforcing conservation invariants and reverting invalid requests.
\textbf{AccumulatorVerifier} provides constant-time membership verification independent of set size, with witnesses maintained off-chain.
\textbf{DIDRegistry} provides auditable identity binding and key rotation.
\textbf{SelectiveDisclosure} represents attribute sets using Merkle root commitments with two-layer access gating, ensuring that unauthorized entities cannot reach the verification path.

\section{Data-Driven Workload Construction and Implementation}

\subsection{Data Sources and Processing}

All experimental data are obtained from publicly available sources and serve only as statistical constraints for synthetic workload construction, not as direct inputs during system execution. Energy trading constraints are derived from PJM day-ahead LMP data \cite{pjm_lmp_data} and EIA 930 hourly operating data \cite{eia_930_data}; network topology and market clearing logic are excluded \cite{saxena_energy_field,vishwakarma_p2p_energy}. Carbon asset constraints are derived from EU ETS aggregated statistics \cite{eu_ets_data,zhang_carbon_shanghai}, with price ranges from ICAP time series \cite{icap_price_data}. All identifiers are anonymously generated; synthetic records do not retain one-to-one correspondence with source data.

\subsection{Synthetic Workload Definition}

The synthetic workload includes only fields required for system evaluation. Energy trading records consist of timestamp, participant identifier, transaction type, energy quantity, settlement price, and region label. Carbon asset records include asset identifier, asset type, credit amount, issuance year, and lifecycle state. Batch processing and capacity constraint models are introduced: during peak hours, higher arrival rates shorten batching windows, increasing amortized cost. When cumulative volume exceeds a daily capacity threshold, a progressive penalty mechanism simulates non-linear cost growth under congestion.

\subsection{Implementation and Reproducibility}

All off-chain processing and workload generation use fixed random seeds, ensuring that identical inputs produce identical results. Experiments are conducted in a local Hardhat environment with fixed compiler versions and account configurations. Source code is available at: https://github.com/xiaohou521/OCAV.

\section{Evaluation}

\subsection{Experimental Setup}

\subsubsection{Experimental Objectives.}

The evaluation focuses on four objectives: (1) whether off-chain records can be independently replayed and field-level tampering deterministically detected; (2) the on-chain gas cost comparison between direct submission and digest commitment; (3) whether lifecycle conservation of carbon assets is enforced by on-chain logic; and (4) the cost of constant-time accumulator verification and selective disclosure.

\subsubsection{Environment and Parameters.}

All experiments are executed in a local Hardhat environment (Solidity 0.8.20, optimizer enabled with 200 runs). The evaluation uses a single-node setup with deterministic block timestamps and fixed account configurations. Synthetic workloads are generated with fixed random seeds to ensure full reproducibility. Energy trading records use PJM-derived price distributions (mean \$35.2/MWh, standard deviation \$12.8/MWh) with hourly arrival rates ranging from 20 to 180 transactions per hour over a 24-hour cycle. Carbon asset records include 50 registered assets with lifecycle states sampled from EU ETS statistics. Batch sizes range from 1 to 128 transactions. The capacity threshold for congestion penalty activation is set at 2{,}000 cumulative daily transactions.

\subsubsection{Evaluation Metrics.}

Security is measured by tamper detection effectiveness. Cost is measured by on-chain gas consumption and its variation. Availability is measured by the rejection rate of invalid operations. Privacy is measured by identity verification effectiveness.

\begin{table}[t]
\centering
\caption{Evaluation metrics and corresponding experiments}\label{tab1}
\begin{tabular}{lll}
\hline
Category & Metric & Experiment \\
\hline
Security & Tamper Detection Effectiveness & Exp1 \\
Cost & Gas Trend, Efficiency, Stability & Exp2/4 \\
Availability & Violation Rejection Effectiveness & Exp3 \\
Privacy & Identity Verification Effectiveness & Exp5 \\
\hline
\end{tabular}
\end{table}

\subsection{Experimental Results}

\subsubsection{Exp1: Settlement Consistency and Replayable Audit.}

This experiment evaluates whether on-chain settlement digests serve as stable audit anchors. Two hundred energy trading settlement digests are generated under PJM price constraints and committed on-chain. Auditors recompute the digests from the corresponding off-chain records using public rules and compare them with the on-chain values. Controlled tampering is applied to six input field categories, with thirty trials per category.

All settlement records pass replay verification in the absence of tampering, yielding a reproducibility rate of 200/200. All 180 tampering attempts are detected as mismatches, with no false negatives observed. These results confirm that deterministic settlement digests provide stable and reproducible audit anchors.

\subsubsection{Exp2: On-Chain Cost Efficiency and Scalability Analysis.}

\begin{figure}[t]
\centering
\includegraphics[width=\linewidth]{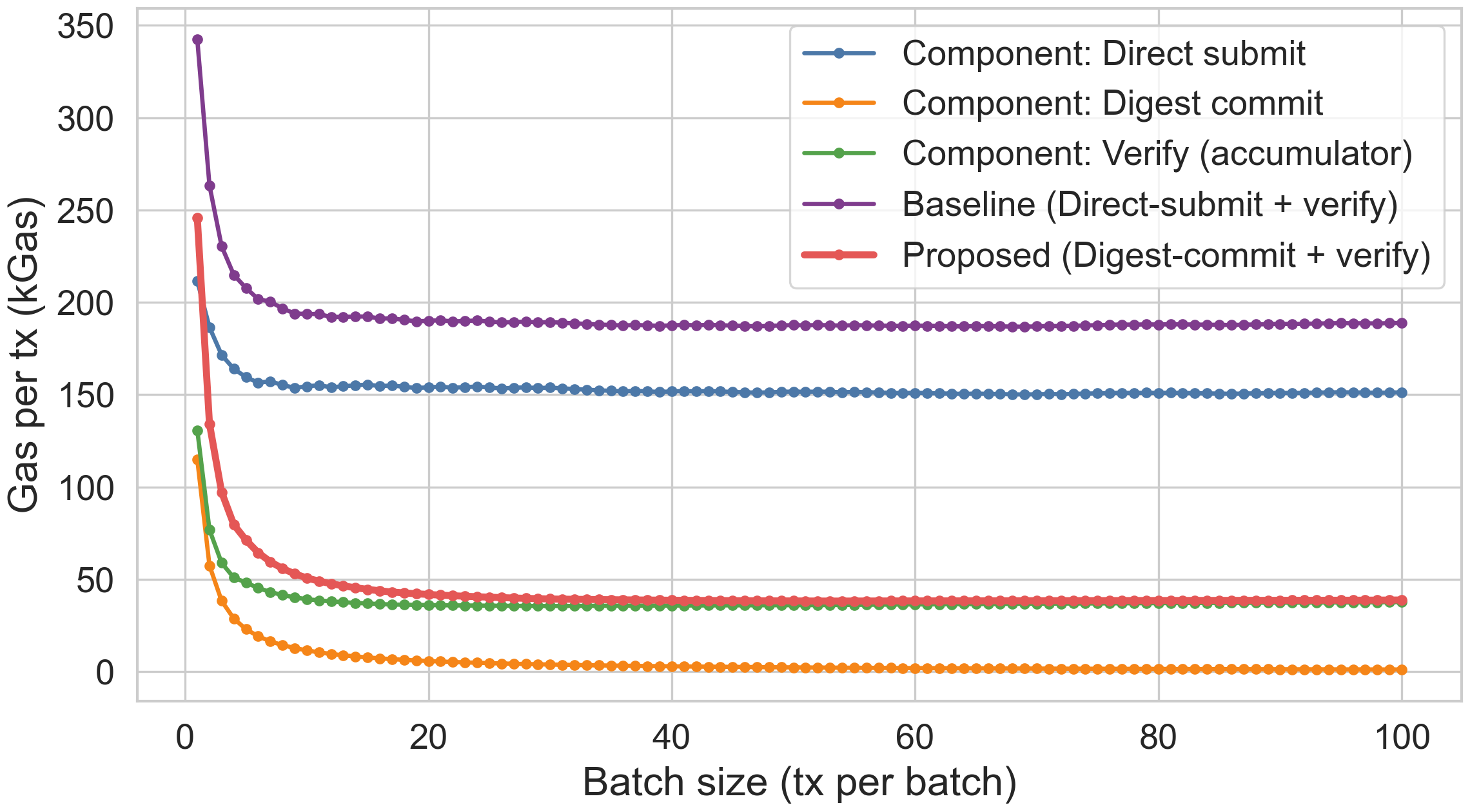}
\caption{Amortized gas per transaction versus batch size.}
\label{fig:exp3_batch}
\end{figure}

\begin{figure}[t]
\centering
\includegraphics[width=\linewidth]{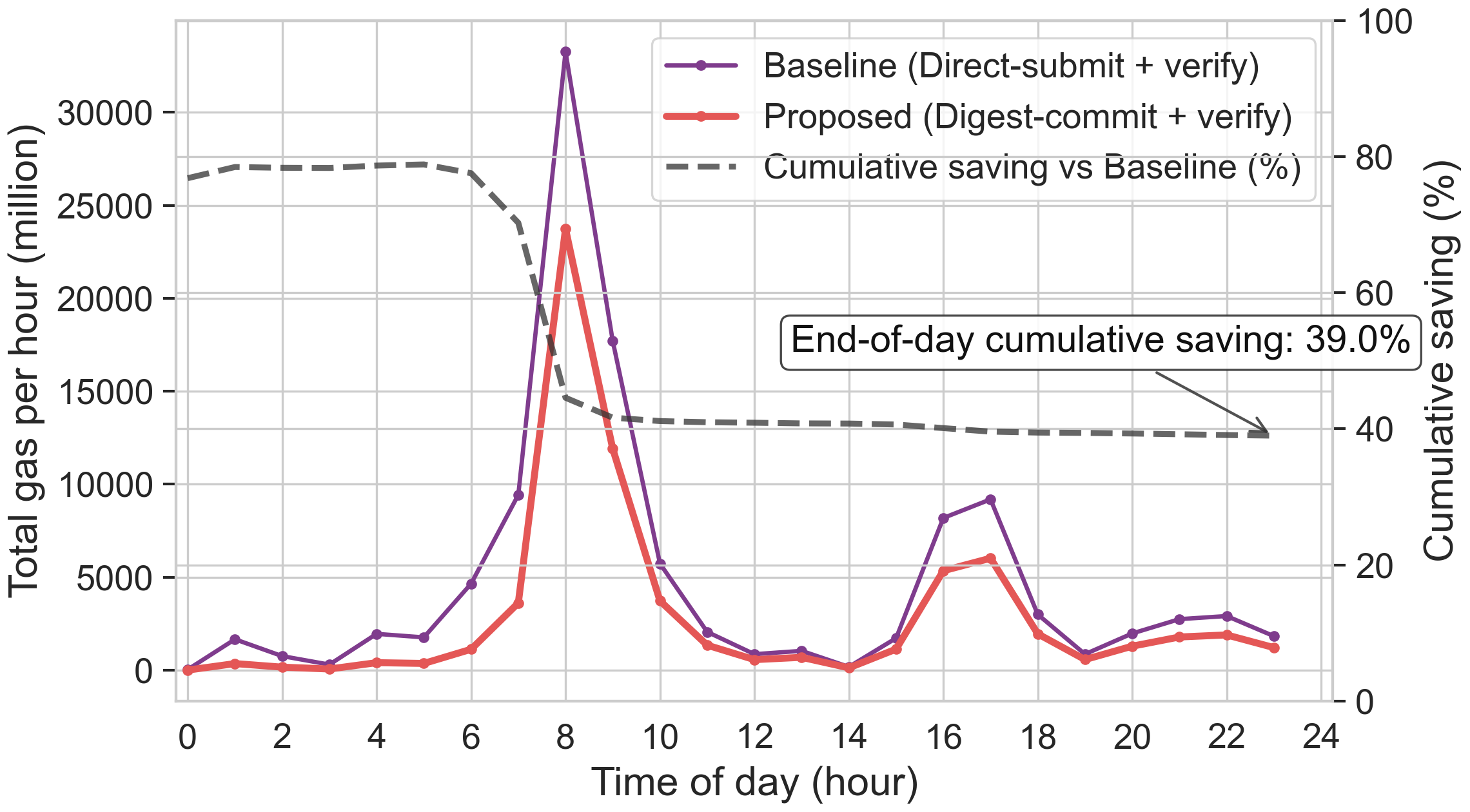}
\caption{Hourly total gas consumption over twenty-four hours.}
\label{fig:exp3_hourly}
\end{figure}

\begin{figure}[t]
\centering
\includegraphics[width=\linewidth]{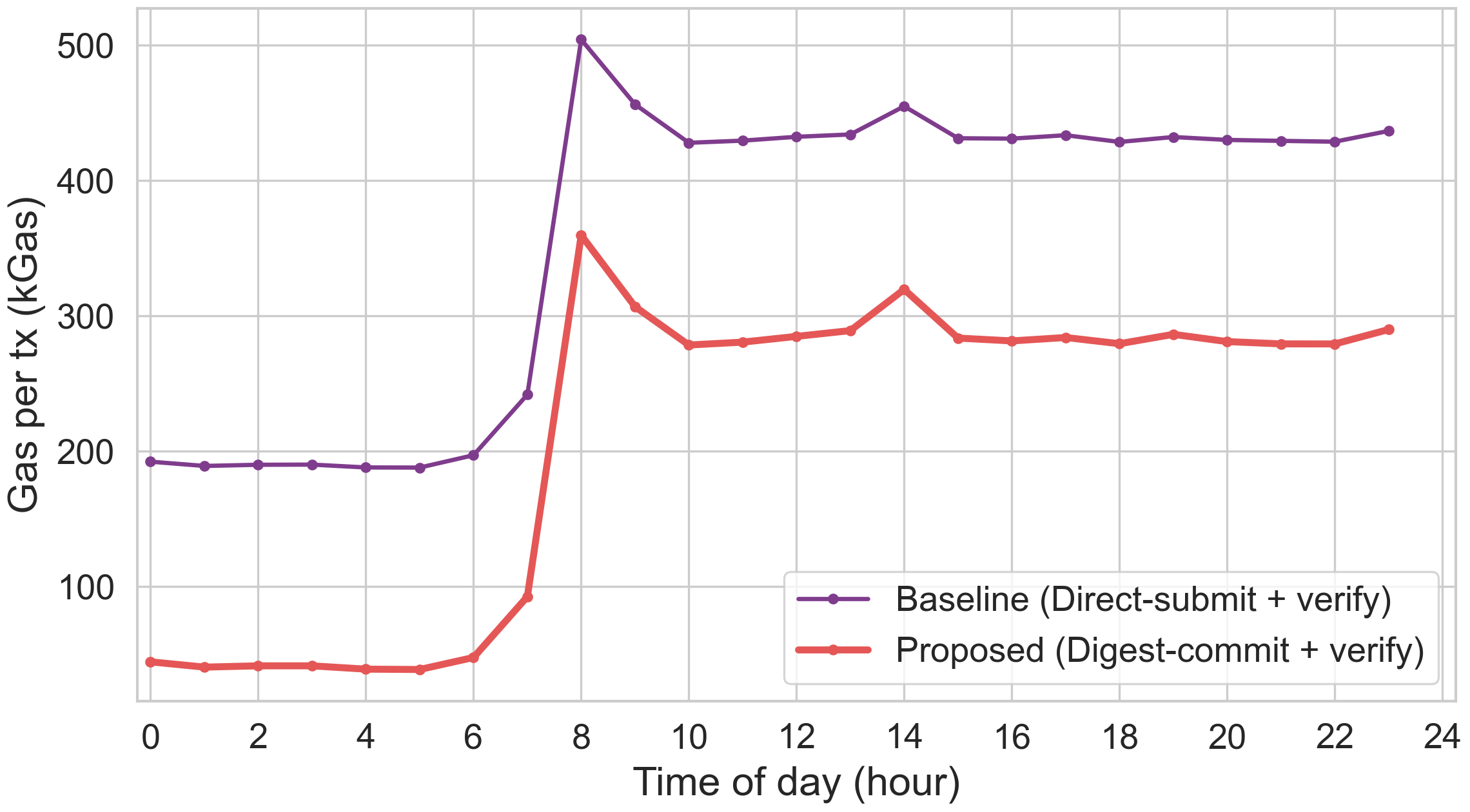}
\caption{Gas per transaction over twenty-four hours under capacity constraints.}
\label{fig:exp3_gptx}
\end{figure}

Two schemes are compared: the Baseline (direct on-chain submission plus verification) and the Proposed (digest commitment plus verification). As shown in Fig.~\ref{fig:exp3_batch}, gas per transaction decreases with batch size due to fixed overhead amortization, and the Proposed scheme outperforms the Baseline across all batch sizes. Fig.~\ref{fig:exp3_hourly} reports hourly gas over a daily cycle; the Proposed scheme remains consistently below the Baseline, achieving a cumulative gas reduction of approximately thirty-nine percent. Fig.~\ref{fig:exp3_gptx} shows that even under capacity constraints with progressive penalty activation, the Proposed scheme maintains lower per-transaction cost.

\subsubsection{Exp3: Carbon Asset Lifecycle Consistency.}

This experiment evaluates whether lifecycle conservation constraints of carbon assets are enforced directly by on-chain logic. Multiple assets are registered, followed by valid transfer and retirement operations. Invalid scenarios are constructed, including over-transfer, over-retirement, and unauthorized actions.

All valid operations preserve lifecycle conservation conditions, and all 30 invalid operations are rejected by the contract. These results indicate that carbon asset conservation is suitable as a globally enforced on-chain invariant.

\subsubsection{Exp4: Constant-Time Accumulator Verification.}

This experiment evaluates the on-chain execution cost of accumulator verification under different set sizes. Verification is performed repeatedly with 10, 50, and 100 elements. The observed gas variation remains below one percent as the set size increases, confirming the constant-time property of the on-chain verification path. Only on-chain verification cost is measured; off-chain witness generation time is excluded.

\subsubsection{Exp5: Identity-Gated Selective Disclosure.}

The system applies a two-layer gating structure: requesters must complete identity authentication via \textbf{DIDRegistry} before attribute-level verification by \textbf{SelectiveDisclosure}. All unauthorized access attempts are rejected, and identity verification effectiveness reaches 100\%, confirming that privacy is enforced through structural identity gating.

\begin{table}[t]
\centering
\caption{Summary of experimental results}\label{tab2}
\begin{tabular}{ll}
\hline
Metric & Result \\
\hline
Tamper Detection Effectiveness & 100\% (180/180) \\
Gas Reduction & 39.0\% (Ours vs Baseline) \\
Violation Rejection Effectiveness & 100\% (30/30) \\
Accumulator Verification Gas Variance & less than 1\% \\
Identity Verification Effectiveness & 100\% \\
\hline
\end{tabular}
\end{table}

Overall, the results confirm that the proposed system meets its design objectives: deterministic digests enable reliable tamper detection, the hybrid architecture reduces on-chain cost by 39\% while preserving auditability, lifecycle constraints prevent irreversible inconsistencies, and identity-gated verification ensures structural privacy.

\section{Discussion and Limitations}

This work focuses on settlement and audit infrastructure rather than modeling energy markets or carbon trading policies. The experiments do not reproduce real-world market clearing rules or pricing mechanisms; public data are used only to constrain time scales and magnitudes. All performance metrics are restricted to on-chain execution paths. The lifecycle consistency model enforces quantity conservation as a minimal condition and is not equivalent to official measurement, reporting, and verification procedures.

\section{Conclusion}

This paper presents a hybrid on-chain and off-chain settlement and audit framework for distributed energy trading and carbon asset management. Through deterministic settlement digests and replayable auditing, the framework preserves core immutability while significantly reducing on-chain storage and execution overhead. Experiments demonstrate settlement consistency verification, tamper detection, and lifecycle conservation without additional trust assumptions. The results indicate that verifiable hybrid settlement provides a practical design path for audit-intensive and high-frequency infrastructure systems.

\begin{credits}
\subsubsection{\ackname}
This work is supported by the Open Research Project of the State Key Laboratory of Industrial Control Technology, China (Grant No. ICT2025B70). Supported by Open Fund of Advanced Cryptography and System Security Key Laboratory of Sichuan Province (Grant No. SKLACSS-202303). Supported by Jiangxi Provincial Natural Science Foundation (Grant No. 20242BAB20041).

\subsubsection{\discintname}
The authors have no competing interests to declare that are relevant to the content of this article.
\end{credits}
\bibliographystyle{splncs04}
\bibliography{references}
\end{document}